# Defining the payback period for nonconventional cash flows: an axiomatic approach

Mikhail V. Sokolov[a,b,c,d]

[a] European University at St. Petersburg, 6/1A Gagarinskaya st., St. Petersburg, 191187, Russia
[b] Centre for Econometrics and Business Analytics (CEBA). St. Petersburg State University, 7/9 Universitetskaya nab., St. Petersburg, 199034, Russia
[c] Institute for Regional Economic Studies RAS, 38 Serpukhovskaya st., St. Petersburg, 190013, Russia
[d] HSE University, 16 Soyuza Pechatnikov st., St. Petersburg, 190121, Russia

**Abstract** The payback period is unambiguously defined for conventional investment projects—projects in which a series of cash outflows is followed by a series of cash inflows. Its definition for nonconventional projects is more challenging, since their balances (cumulative cash flow streams) may have multiple break-even points. Academics and practitioners offer a few contradictory recipes to manage this issue, suggesting to use the first break-even point of the balance, the last break-even point of the balance, or the moment in time at which the cumulative sum of net cash inflows first exceeds the total sum of net cash outflows. In this paper, we show that the last break-even point of the project balance is the only definition of the payback period consistent with a set of economically meaningful axioms. An analogous result is established for the discounted payback period.

---

E-mail address: mvsokolov@eu.spb.ru
The paper is prepared based on the outcomes of research under the assignment of IRES RAS on the topic “Comprehensive study of demographic and socio-economic processes in the context of social development turbulence using demographic, economic-mathematical, and econometric methods at the macro and regional levels”. Project No. 125011300216-4.

## 1. Introduction

The payback period remains one of the most widely utilized capital budgeting techniques in managerial practice, consistently ranking among the most popular methods for evaluation of investment performance (Siziba & Hall, 2021; Graham, 2022). Its enduring appeal lies in its ease of computation and straightforward interpretation. Despite its practical prevalence, the payback period is almost universally criticized in the academic finance literature for its significant theoretical shortcomings (Weingartner, 1969; Vilensky et al., 2002, pp. 320–321; Fabozzi & Peterson, 2003, p. 404; Brealey et al., 2020, p. 113), primarily its disregard for the time value of money and all cash flows beyond the payback cutoff. There are, however, a few formal arguments supporting its use as an investment decision criterion under risk (Wambach, 2000; Boyle & Graeme, 2006) and information asymmetry (Narayanan, 1985; Jahnke & Simons, 2008).

Traditional treatments of the payback period method as well as of its discounted counterpart focus almost exclusively on conventional projects (Rappaport, 1965; Liao, 1976; Bhandari, 1985, 2009; Lefley, 1996; Fabozzi & Peterson, 2003, pp. 402–410; Brealey et al., 2020, Section 5-2), projects in which a series of cash outflows is followed by a series of cash inflows. For such projects, the payback period is unambiguously defined as a break-even point of the project balance (cumulative cash flow stream), providing a clear metric for evaluating capital recovery and liquidity. This restrictive focus leaves a significant gap in the method's applicability, as many real-world investment opportunities exhibit nonconventional cash flow patterns, for which the definition of payback period is more challenging due to multiple break-even points. Scarce academic and practitioner literature managing this issue offers a few contradictory recipes, suggesting to define the payback period as the first break-even point of the project balance, the last break-even point of the project balance, or the moment in time at which the cumulative sum of net cash inflows first exceeds the total sum of net cash outflows. The first recipe—the first break-even point of the project balance—is mainly used in stochastic settings, where the payback period is often modeled as the first passage time: the time at which the stochastic cumulative cash flow first reaches zero (Weingartner, 1969; Kim et al., 2013). The last break-even point of the project balance (more precisely, the shortest time period after which the balance becomes and stays nonnegative) seems to be the most widely used modification of the conventional payback period (Hajdasiński, 1993; Vilensky et al., 2002, Section 8.3), not least because of its clear economic intuition: capital is fully recovered only at the time from which the cumulative cash flow remains nonnegative. The metric representing the moment in time at which the cumulative sum of net cash inflows first exceeds the total sum of net cash outflows is discussed, e.g., in Nijhawan (2012, p. 170). See also Cheremushkin (2016) for a comparative study of the latter two metrics.

In this paper, we show that the last break-even point of the project balance is the only extension of the conventional definition of payback period consistent with a set of economically meaningful axioms. A similar result is established for the discounted payback period. Closest to our work is Sokolov (2024, Section 4.2), who considered a more restrictive set of axioms and proved an impossibility result that every extension of the conventional definition of payback period (defined as a unique break-even point of the project balance) to a larger domain necessarily violates an axiom from that set. Relaxing some of those axioms, in particular, the requirement that an upper contour set of the payback period is closed under addition (note that the dual requirement for a lower contour set is natural: it guarantees that for each given maximum acceptable payback period, the union of economically acceptable projects is economically acceptable), we obtain a positive result.

The paper is organized as follows. Section 2 contains preliminaries and introduces the space of cash flow streams we deal with. In Section 3, our main result—an axiomatic characterization of the payback period—is presented. In Section 4, the proposed approach is adopted to axiomatize the discounted payback period. The paper primarily deals with investment projects with finite lifetimes, identifying them with the change in the cash balance they generate. Two more general setups are outlined in Section 5. All proofs are given in the Appendix.

## 2. Investment projects

We begin with basic definitions and notation. $\mathrm{R}_{++}$, $\mathrm{R}_{+}$, $\overline{\mathrm{R}}_{+} = \mathrm{R}_{+} \cup \{+\infty\}$, and $\mathrm{R}$ are the sets of positive, nonnegative, extended nonnegative, and all real numbers, respectively. For any $\tau \in \mathrm{R}_{+}$, let $1_{\tau}$ be the function on $\mathrm{R}_{+}$ given by

$$1_{\tau}(t) = \begin{cases} 1, & t \geq \tau \\ 0, & t < \tau \end{cases}.$$

Throughout the paper an investment project is identified with the cash flow stream it generates. Following Norberg (1990), Promislow (1994), Smolyak (2002), and Armerin (2014), we prefer to describe a project by means of cumulative (rather than net) cash flow stream. The main advantage of this setup is that it enables a uniform treatment of the two standard settings—discrete (a setting in which a project is represented by means of a sequence of pairs $(c_k, \tau_k)$, $k = 0,1, \ldots$, where $c_k$ is the net cash flow at time $\tau_k$) and continuous (a setting in which a project is modeled via an instantaneous rate of payment). We restrict our analysis to projects with deterministic cash flows and finite lifetimes (for a discussion of more general setups, see Section 5). Specifically, by a (*investment*) *project* we mean a function $x: \mathrm{R}_{+} \to \mathrm{R}$ that satisfies the following three conditions:

(A) $x$ is of locally bounded variation (i.e., it has bounded variation on each interval $[0, t]$, $t \in \mathrm{R}_{++}$);
(B) there is $T \in \mathrm{R}_{+}$ such that $x$ is constant on $[T, +\infty)$;
(C) $x$ is right-continuous.

The function $x$ in the definition is interpreted as the cumulative (deterministic) cash flow stream the project generates, i.e., $x(t)$ is the balance of the project at time $t$ (the difference between cumulative cash inflows and cumulative cash outflows over the time interval $[0, t]$ with a standard normalization that $t = 0$ denotes the present moment). In view of the Jordan decomposition theorem (Monteiro et al., 2018, Theorem 2.1.21, p. 17), condition (A) states that a project $x$ can be represented in the form $x = x_{+} - x_{-}$, where $x_{+}$ and $x_{-}$ are nonnegative nondecreasing functions. Such a representation is vital for $x$ to be interpreted as a cumulative cash flow stream as, by definition, it is the net of cumulative cash inflow and outflow streams, which are nonnegative nondecreasing functions of time. Condition (B) states that a project has a finite lifetime. Conditions (A) and (B) imply that $x$ is of bounded variation, and in particular bounded (Monteiro et al. 2018, Remark 2.1.9, p. 12), imposing realistic bounds on projected cash flows. Finally, condition (C) properly incorporates discrete cash flow setup. Taken together, conditions (A)–(C) imply that a project can be decomposed into the sum of a step function and a continuous function (Monteiro et al., 2018, Theorem 2.6.1, p. 35), representing, respectively, the discrete and continuous parts of the cumulative cash flow stream the project generates. Explicitly, the discrete part is a function on $\mathrm{R}_{+}$ with a representation $\sum_{k=0}^{\infty} c_k 1_{\tau_k}$ for some sequences $\{\tau_k\} \subset \mathrm{R}_{+}$ and $\{c_k\} \subset \mathrm{R}$ with $\sum_{k=0}^{\infty} |c_k| < +\infty$. Here $c_k$ is interpreted as the net cash flow at time $\tau_k$. And the continuous part is a continuous function on $\mathrm{R}_{+}$ of bounded variation. An important example of a continuous part is given by $t \mapsto \int_0^t c(\tau) \mathrm{d}\tau$, where $c$ is an integrable function, whose value $c(\tau)$ is interpreted as the instantaneous net cash flow (the instantaneous rate of payment) at time $\tau$. We let $\mathrm{P}$ denote the vector space of all projects endowed with a topology. The zero project—the zero function from $\mathrm{R}_+$ into $\mathrm{R}$—is denoted by $\mathbf{0}$. We write $x \preccurlyeq y$ if $x(t) \le y(t)$ $\forall t \in \mathrm{R}_+$.

## 3. The payback period: an axiomatic approach

We begin with a concise overview of (the most conservative variant of) the conventional definition of payback period and its limitations. Set $\mathrm{C} = \{x \in \mathrm{P}$: there exists $\tau \in \mathrm{R}_{++}$ such that $x$ is negative on $[0, \tau)$ and nonnegative on $[\tau, +\infty)\}$. The *conventional payback period* (*CPP*, for short) is the function $CPP\colon \mathrm{C} \to \mathrm{R}_{++}$ that maps each $x \in \mathrm{C}$ to the value $\tau$ that appears in the definition of the set $\mathrm{C}$ (note that $\tau$ is unique).[1] Put differently, the CPP is a break-even point of the project balance, provided that such a point exists and is unique. Clearly, $\mathrm{C}$ contains as a proper subset all conventional projects (projects, in which a series of cash outflows is followed by a series of cash inflows) that pay off; that is, the CPP is well defined for those projects. On the other hand, it follows from the definition that CPP is undefined for projects that have multiple sign changes in their cumulative cash flow streams. To illustrate, consider the project $z$ which generates the net cash flow $-12$ monetary units at time $0$, $13$ at time $1$, $-2$ at time $2$, and $8$ at time $3$. Its cumulative cash flow stream, $z = -12 \cdot 1_0 + 13 \cdot 1_1 - 2 \cdot 1_2 + 8 \cdot 1_3$, has two break-even points—at time $1$ and $3$. Therefore, $CPP(z)$ is undefined. CPP is also undefined for projects with negative cumulative cash flow streams (as well as nonnegative ones), which means that they never pay off. More importantly, even if projects $x_1, \dots, x_n$ have CPPs, the corresponding pool of projects, $x_1 + \dots + x_n$, may have no CPP, since $\mathrm{C}$ is not closed under addition. To illustrate the problem, consider an investor who screens projects using a maximum acceptable payback cutoff $d$, expecting that the CPP for the pool of implemented projects will also meet the target. Indeed, this holds true as $CPP(x_1) \le d$ & … & $CPP(x_n) \le d \Rightarrow CPP(x_1 + \dots + x_n) \le d$, whenever $x_1, \dots, x_n, x_1 + \dots + x_n \in \mathrm{C}$. But the problem is that $x_1, \dots, x_n \in \mathrm{C} \nRightarrow x_1 + \dots + x_n \in \mathrm{C}$, i.e., in general, the CPP of the pool of projects is undefined. For instance, the projects $u = -10 \cdot 1_0 + 15 \cdot 1_1$ and $v = -2 \cdot 1_0 - 2 \cdot 1_1 - 2 \cdot 1_2 + 8 \cdot 1_3$ have the CPPs of 1 and 3, respectively. However, as $u + v = z$, $CPP(u + v)$ is undefined.

We proceed with an axiomatic definition of payback period. Let $\mathrm{P}_0 = \{-a1_0 + b1_\tau\colon 0 < a \le b, \tau > 0\}$ be the set of investment projects with two transactions—an initial outlay and a final inflow—that pay off and let $CPP_0\colon \mathrm{P}_0 \to \mathrm{R}_{++}$ be given by $CPP_0(-a1_0 + b1_\tau) = \tau$ (i.e., $CPP_0$ is the restriction of $CPP$ to $\mathrm{P}_0$). We say that a function $D\colon \mathrm{P} \to \overline{\mathrm{R}}_+$ is a *payback period* if the following three conditions hold.

*Compliance* (*COMP*): $x \in \mathrm{P}_0 \Rightarrow D(x) = CPP_0(x)$.
*Monotonicity* (*MON*): $x \preccurlyeq y \Rightarrow D(x) \ge D(y)$.
*Aggregation consistency* (*ACONS*): $D(x + y) \le \max\{D(x), D(y)\}$.

The concept of payback period is unambiguously defined for investment projects from $\mathrm{P}_0$ (for instance, the payback period of a $\tau$-year deposit at a nonnegative interest rate, with interest paid at maturity, is $\tau$). Condition COMP requires $D$ to comply with this definition. Note that we do not ask for a payback period to agree with $CPP$ on $\mathrm{C}$, which is a stronger condition. The motivation is that the literature knows payback metrics that differ from the conventional one on $\mathsf{C}$. For instance, the payback metric mentioned in Section 1 and representing the moment in time at which the cumulative sum of net cash inflows first exceeds the total sum of net cash outflows does not reduce to $CPP$ on $\mathsf{C}$ (Cheremushkin, 2016). Most, if not all, known payback metrics, however, reduce to $CPP_0$ on $\mathsf{P}_0$.

---

[1] In what follows, we distinguish between CPP (the abbreviation for the term) and $CPP$ (the function from $\mathrm{C}$ to $\mathrm{R}_{++}$ that maps a project to the value $\tau$ that appears in the definition of the set $\mathrm{C}$.

Condition MON ensures that a higher balance provides a lower payback period. This condition combines both the usual "the more money, the better" assumption and liquidity prioritization. In particular, accelerating cash inflows and delaying cash outflows reduce a payback period (note that $1_t \preccurlyeq 1_\tau$ and $-1_\tau \preccurlyeq -1_t$, whenever $t > \tau$). One can prove that under COMP and ACONS, condition MON is equivalent to the following one: $\mathbf{0} \preccurlyeq x \Rightarrow D(x) = 0$. That is, projects with nonnegative balance must have zero payback period.

Condition ACONS relates payback period for a pool of projects with payback periods of its components. Decision makers often specify a maximum acceptable payback period (MAPP); they consider a project to be economically acceptable if it is paid back within the MAPP.[2] Condition ACONS asserts that for each given MAPP, the union of economically acceptable projects is economically acceptable. Indeed, ACONS can be equivalently stated as follows: for any $d \in \mathsf{R}_{++}$, $D(x) \leq d$ & $D(y) \leq d \Rightarrow D(x + y) \leq d$. This condition validates the following natural guidance: to guarantee a target payback period for a pool of projects it suffices to keep the target for each project in the pool.

A cash flow stream contains future components that are often measured with error. This requires investment decisions to be robust to minor perturbations of cash flows. In relation to accept/reject investment decisions based on a payback period, a reasonable robustness requirement (which we do not consider part of the definition of a payback period) can be stated as follows.

*Lower semicontinuity* (*LSC*): for any $d \in \mathsf{R}_{++}$, the set $\{x \in \mathsf{P}: D(x) > d\}$ is open.

Condition LSC asserts that for each given MAPP, minor perturbations of a cash flow stream cannot convert an economically unacceptable project (i.e., a project whose payback period is longer than the MAPP) into an economically acceptable one. LSC can also be viewed as a non-manipulability condition. Indeed, consider a grantor who exclusively funds projects with a payback period that does not exceed a MAPP $d$. If the set $\{x \in \mathsf{P}: D(x) > d\}$ is not open, an ineligible project could be rendered eligible by an arbitrarily small perturbation, creating a vulnerability that an unscrupulous grantee might exploit. Although condition LSC is not included in the definition of a payback period, it can be deduced from COMP, MON, and ACONS under fairly weak conditions on the topology on $\mathsf{P}$, as shown in the proposition below.

A more restrictive set of axioms including COMP, MON, ACONS, and LSC is considered in Sokolov (2024, Section 4.2). The key difference is that in addition to ACONS, he also imposes its dual: $\min\{D(x), D(y)\} \leq D(x + y)$. The latter axiom is essential in the context of profitability and rate of return measurement (Promislow, 1997; Vilensky & Smolyak, 1999; Sokolov, 2023, 2024), meaning that for each given hurdle rate, the union of profitable projects (in the sense that their rates of return are at least the hurdle rate) is profitable. However, this axiom appears less intuitive in the context of payback metrics as a decision maker usually bounds them from above rather than from below. The main finding of Sokolov (2024, Section 4.2) is an impossibility result that $CPP$ cannot be extended to a larger domain consistent with the set of axioms he considered. By relaxing the requirement for the dual of ACONS, we obtain a positive result.

[2] Note that determining the MAPP presents a distinct challenge (e.g., see Yard, 2000).

**Proposition 1.**

1°. *Conditions COMP, MON, and ACONS are independent, i.e., any two of them do not imply the third.*

2°. *COMP, MON, and ACONS ⇒ LSC, provided that the topology on* P *is finer than the topology of pointwise convergence.*

3°. *For a function $D\colon \mathrm{P} \to \overline{\mathrm{R}}_+$, the following statements are equivalent:*

(a) *$D$ is a payback period;*

(b) $D(x) = \inf\{\tau \in \mathrm{R}_+ : x(t) \geq 0 \ \ \forall t \geq \tau\}$;[3]

(c) $D(x) = \inf\{CPP_0(y) : y \in \mathrm{P}_0, y \preccurlyeq x\}$.

It follows from part 3° of Proposition 1 that a payback period is unique. In view of this, in what follows, we call the function in representation (b) *the* payback period and denote it by $PP$. $PP$ maps a project to the minimum time period $\tau$ (if any) such that the balance $x(t)$ of the project remains nonnegative for all $t \geq \tau$. If there is no such $\tau$, i.e., the project never pays off, then the payback period is equal to $+\infty$. This formulation is precisely the definition of payback period advocated in Hajdasiński (1993) and Vilensky et al. (2002, Section 8.3). Clearly, $PP$ reduces to $CPP$ on C. $PP$ is positively homogeneous of degree zero, that is, it takes no account of the investment size and hence is a relative measure. By representation (c), the payback period of a project $x$ has an intuitive interpretation: it is the greatest lower bound of the payback periods of projects with two transactions dominated by $x$. One can show that the infimum in part (b) (resp. (c)) is attained, provided that $D(x) \in \mathrm{R}_+$ (resp. $D(x) \in \mathrm{R}_{++}$). In particular, part (b) (resp. (c)) provides the following characterization of the set of economically acceptable projects: for a given MAPP $d \in \mathrm{R}_{++}$, a project $x \in \mathrm{P}$ is economically acceptable, i.e., the payback period of $x$ does not exceed $d$, if and only if $x$ is nonnegative on $[d, +\infty)$ (resp. there exists $y \in \mathrm{P}_0$ such that $y \preccurlyeq x$ and $CPP_0(y) = d$). It follows from the proof that Proposition 1 remains valid if the set P is replaced by the set of all bounded right-continuous real-valued functions on $\mathrm{R}_+$. By part 1°, none of the three axioms can be omitted without altering the result.

Part 2° of Proposition 1 shows that for the payback period, LSC holds conditional on topology chosen. One can also deduce from 2° that $PP$ satisfies the following "topology-free" variant of the robustness requirement, which we refer to as algebraic lower semicontinuity: for each given MAPP $d \in \mathrm{R}_{++}$, the set $\{x \in \mathrm{P} : PP(x) > d\}$ of economically unacceptable projects is algebraically open, that is, for any $x$ with $PP(x) > d$ and $y \in \mathrm{P}$, there is $\varepsilon > 0$ such that $PP(x + ty) > d \ \forall t \in (0, \varepsilon)$.[4]

Turning back to the examples of projects given at the beginning of this section, we have $PP(u) = CPP(u) = 1$, $PP(v) = CPP(v) = 3$, and $PP(z) = 3$. Note that $PP(u + v) = PP(z) \leq \max\{PP(u), PP(v)\}$. As a result, for each given MAPP, if $u$ and $v$ are economically acceptable, then so is $u + v$.

Proposition 1 demonstrates that any suggestion for the definition of payback period different from $PP$ necessarily violates an economically meaningful axiom. In particular, as for the two metrics mentioned in Section 1, one can show that the first break-even point of the project balance does not satisfy ACONS, whereas the moment in time at which the cumulative sum of net cash inflows first exceeds the total sum of net cash outflows does not satisfy MON (however, it satisfies a weaker monotonicity condition—the higher the net cash flow, the lower the metric). This means, e.g., that a decision maker must be careful when using the first break-even point of the project balance to set a target payback period in the presence of several independent investment proposals: implementation of those proposals that meet the target does not guarantee the resulting pool of proposals to achieve the target.

[3] Here and hereafter, we use the convention $\inf \emptyset = +\infty$.

[4] If we endow P with the topology of pointwise convergence, then for any given MAPP, the set of economically unacceptable projects is open (by 2°) and, hence, algebraically open.

## 4. The discounted payback period

In this section, the axiomatic approach we propose is adopted to formalize the discounted payback period.

We begin with preliminary definitions, introducing discounting and cumulative discounted cash flow streams. A continuous function $\alpha\colon \mathrm{R}_+ \to \mathrm{R}_{++}$ satisfying $\alpha(0) = 1$ is called a *discount function*. As usual, $\alpha(t)$ is interpreted as the discount factor at time $t$ (the present worth of receiving a monetary unit at time $t$). We allow $\alpha(t)$ to be greater than 1 as some individual and institutional investors may have preferences for future payoffs (e.g., central banks of some countries practice negative interest rates). For a project $x$ and a discount function $\alpha$, set $x^{(\alpha)}(t) = x(0) + \int_0^t \alpha(\tau)\mathrm{d}x(\tau)$, $t \in \mathrm{R}_+$, where the integral is the Riemann-Stieltjes integral.[5] The function $x^{(\alpha)}$ represents the cumulative discounted cash flow stream associated with $x$ and $\lim_{t\to+\infty} x^{(\alpha)}(t)$ is the net present value of the project $x$. For a discrete cash flow stream $x = \sum_{k=0}^{\infty} c_k 1_{\tau_k}$, $x^{(\alpha)}$ reduces to the discounted sum $x^{(\alpha)} = \sum_{k=0}^{\infty} \alpha(\tau_k) c_k 1_{\tau_k}$. For a continuously differentiable cumulative cash flow stream $x$ with $x(0) = 0$, $x^{(\alpha)}$ reduces to another familiar expression appeared in continuous-time setup, $x^{(\alpha)}(t) = \int_0^t \alpha(\tau)x'(\tau)\mathrm{d}\tau$, where $x'(\tau)$ is interpreted as the instantaneous net cash flow (the instantaneous rate of payment) at time $\tau$. Note that for any discount function $\alpha$, the map $x \longmapsto x^{(\alpha)}$ is a bijection from P onto P (Monteiro et al., 2018, Corollary 6.5.5, p. 172); the inverse map is given by $x \longmapsto x^{(1/\alpha)}$.

The discounted payback period is a modification of the conventional one that incorporates the time value of money. Given a discount function $\alpha$, the *conventional discounted payback period*, in the most conservative variant, is the function $CDPP^{(\alpha)}$ defined on the set $\{x \in \mathrm{P}\colon x^{(\alpha)} \in \mathrm{C}\}$ by $CDPP^{(\alpha)}(x) = CPP(x^{(\alpha)})$. It can be shown (see Lemma 1 in the Appendix) that if a discount function $\alpha$ is nonincreasing and $x, x^{(\alpha)} \in \mathrm{C}$, then $CPP(x) \le CDPP^{(\alpha)}(x)$ (this fact is well known in the literature in the special case of an exponential discount function and a conventional cash flow (Bhandari, 1985, footnote 8)). Hence, $CDPP^{(\alpha)}$ is a more conservative payback metric than $CPP$. The conventional discounted payback period inherits the limitations of the CPP mentioned in the previous section.

We proceed with an axiomatic definition of discounted payback period. Given a discount function $\alpha$, we say that a function $D\colon \mathrm{P} \to \overline{\mathrm{R}}_+$ is a *discounted payback period* if it satisfies ACONS and the following two conditions.

*$\alpha$-compliance* ($\alpha$-*COMP*): $x^{(\alpha)} \in \mathrm{P}_0 \Rightarrow D(x) = CPP_0(x^{(\alpha)})$.
*$\alpha$-monotonicity* ($\alpha$-*MON*): $x^{(\alpha)} \preccurlyeq y^{(\alpha)} \Rightarrow D(x) \ge D(y)$.

[5] See Monteiro et al. (2018, Chapter 5) for a review of the Riemann-Stieltjes integral.

Put differently, the definition of a discounted payback period is the one for a payback period with axioms COMP and MON replaced by their discounted counterparts, $\alpha$-COMP and $\alpha$-MON. The interpretation of conditions $\alpha$-COMP and $\alpha$-MON is similar to those of COMP and MON and is omitted. It follows from Lemma 1 in the Appendix that if a discount function $\alpha$ is nonincreasing, then for any $x, y \in \mathrm{P}$, $x \preccurlyeq y \Rightarrow x^{(\alpha)} \preccurlyeq y^{(\alpha)}$. Hence, for a nonincreasing discount function, a discounted payback period satisfies MON.

The following result is a direct consequence of the definitions of a payback period and a discounted payback period.

**Proposition 2.**

*Given a discount function $\alpha$, there is a unique discounted payback period. It is given by $DPP^{(\alpha)}(x) = PP(x^{(\alpha)})$.*

It follows from Proposition 2 that a discounted payback period is unique. It is defined as the minimum time period $\tau$ (if any) such that the discounted balance of the project is nonnegative for all $t \geq \tau$ (if there is no such $\tau$, then the discounted payback period is equal to $+\infty$), the formulation being identical to the discounted payback period definition presented in Hajdasiński (1993) and Vilensky et al. (2002, Section 8.3). One can also deduce from Propositions 1 and 2 that conditions $\alpha$-COMP, $\alpha$-MON, and ACONS are independent and together imply algebraic lower semicontinuity.

Recall that if a discount function $\alpha$ is nonincreasing, then $CPP(x)$ does not exceed $CDPP^{(\alpha)}(x)$, provided that the metrics are well defined. Note that this inequality, in general, no longer holds for $PP(x)$ and $DPP^{(\alpha)}(x)$.

## 5. Two extensions

In the previous sections, we have dealt with investment projects with finite lifetimes and identified them with the (bounded) cumulative cash flow streams they generate. In this section, we outline two more general models of the project space. In subsection 5.1, projects with infinite lifetimes and unbounded cash flows are analyzed. In subsection 5.2, projects are modeled by means of the pair of cumulative cash inflow and outflow streams rather than by their difference—the cumulative cash flow stream. The key finding of this section is that our central result that $PP$ is the only meaningful extension of $CPP$ remains valid in these more general frameworks, albeit at the cost of a minor strengthening of the axioms.

### 5.1. Projects with infinite lifetimes and unbounded cash flows

Some theoretical models operate investment projects with infinite lifetimes and unbounded cumulative cash flow streams (e.g., perpetuity). Condition (B) in our definition of investment project excludes such projects from consideration. In this subsection, we outline how to modify the exposition to incorporate this type of projects. Namely, one can show that Propositions 1 and 2 remain valid, provided that the following three replacements are made in the definition of investment project and the definition of a (discounted) payback period (specifically, conditions COMP and $\alpha$-COMP):

1. The set $\mathrm{P}$ is replaced by the set $\overline{\mathrm{P}}$ of real-valued functions on $\mathrm{R}_+$ satisfying conditions (A) and (C).

2. The set $\mathrm{P}_0$ is replaced by the set $\overline{\mathrm{P}}_0 = \{x \in \overline{\mathrm{P}}: x(0) < 0$, $x$ is nondecreasing, and there is $\tau \in \mathrm{R}_{++}$ such that $x(\tau) \geq 0\}$.

3. The function $CPP_0$ is replaced by the function $\overline{CPP}_0: \overline{\mathrm{P}}_0 \to \mathrm{R}_{++}$ defined by $\overline{CPP}_0(x) = \min\{\tau \in \mathrm{R}_{++}: x(\tau) \geq 0\}$.

Here $\overline{\mathrm{P}}$ is the set of all projects, including those with infinite lifetimes and unbounded cumulative cash flow streams. $\overline{\mathrm{P}}_0$ is the set of conventional projects (projects, in which an initial cash outflow is followed by a series of cash inflows) that pay off and $\overline{CPP}_0$ is the traditional definition of payback period for these projects. The proof of this result is nearly identical to those of Propositions 1 and 2, and requires only a minor adjustment to the argument for the implication "(a)⇒(b)" in part 3° of Proposition 1. We omit the details.

### 5.2. Gross cash inflows and outflows

While net cash flow is a crucial metric for investment appraisal, focusing solely on it and ignoring the underlying gross cash inflows and outflows can potentially obscure several important aspects of an investment project like project scale, timing, liquidity needs, and sensitivity. Note that some capital budgeting metrics (like several variants of the modified internal rate of return (Vilensky et al., 2002, Section 8.2.4) and the benefit-cost ratio (Vilensky et al., 2002, Section 8.2.1; Fabozzi & Peterson, 2003, pp. 416–419)) are functions of gross cash inflows and outflows rather than net cash flow. In this subsection, we model projects by means of the pair of cumulative cash inflow and outflow streams rather than by their difference—the cumulative cash flow stream. The key takeaway from this subsection is that $PP$ continues to be the unique meaningful extension of $CPP$ in this extended framework.

Let X be the set of nondecreasing and right-continuous functions from $\mathrm{R}_+$ to $\mathrm{R}_+$ that are constant on $[T, +\infty)$ for some $T \in \mathrm{R}_+$. A project is represented by a pair $(x_+, x_-) \in \mathrm{X} \times \mathrm{X}$, where $x_+$ and $x_-$ are interpreted respectively as cumulative cash inflow and outflow streams associated with the project. Note that $\mathrm{P} = \mathrm{X} - \mathrm{X}$, i.e., this setup extends the one in Section 2. We equip P with a vector topology which is finer than the topology of pointwise convergence, X with the subspace topology induced by the topology on P, and $\mathrm{X} \times \mathrm{X}$ with the product topology.

Consider the following conditions on a function $D: \mathrm{X} \times \mathrm{X} \to \overline{\mathrm{R}}_+$ (interpreted as a payback metric).

*COMP'*: $x_+ - x_- \in \mathrm{P}_0 \Rightarrow D(x_+, x_-) = CPP_0(x_+ - x_-)$.
*MON'*: $x_+ \preccurlyeq y_+ \ \& \ y_- \preccurlyeq x_- \Rightarrow D(x_+, x_-) \geq D(y_+, y_-)$.
*ACONS'*: $D(x_+ + y_+, x_- + y_-) \leq \max\{D(x_+, x_-), D(y_+, y_-)\}$.
*LSC'*: for any $d \in \mathrm{R}_{++}$, the set $\{(x_+, x_-) \in \mathrm{X} \times X: D(x_+, x_-) > d\}$ is open.

Conditions COMP', MON', and ACONS' are natural counterparts of COMP, MON, and ACONS for payback metrics defined on $\mathrm{X} \times \mathrm{X}$ in the following sense: if $D$ depends on $(x_+, x_-)$ only through $x_+ - x_-$, i.e., $D(x_+, x_-) = f(x_+ - x_-)$ for some $f: \mathrm{P} \to \overline{\mathrm{R}}_+$, then $D$ satisfies COMP' (resp. MON', ACONS') if and only if $f$ satisfies COMP (resp. MON, ACONS). Condition LSC' is weaker than LSC in the sense that if $D$ depends on $(x_+, x_-)$ only through $x_+ - x_-$, i.e., $D(x_+, x_-) = f(x_+ - x_-)$ for some $f: \mathrm{P} \to \overline{\mathrm{R}}_+$, and $f$ satisfies LSC, then $D$ satisfies LSC'.

The following result is the analogue of the key part of Proposition 1.

**Proposition 3.**

*For a function $D\colon \mathrm{X}\times\mathrm{X}\to\overline{\mathrm{R}}_+$, the following statements are equivalent:*

(a) *D satisfies conditions COMP', MON', ACONS', and LSC';*

(b) *D satisfies conditions COMP', MON', ACONS', and the following condition: $x_- \preccurlyeq x_+ \Rightarrow D(x_+, x_-) = 0$;*

(c) $D(x_+, x_-) = PP(x_+ - x_-)$.

Proposition 3 shows that the last break-even point of the project balance is still the only meaningful extension of the CPP to a payback metric defined on $\mathrm{X}\times\mathrm{X}$. However, this broader result comes at the cost of a minor strengthening the axioms. Condition LSC' in part (a) (note that LSC' in Proposition 3 is an assumption, whereas LSC in Proposition 1 is a consequence) as well as the additional monotonicity condition in part (b) cannot be omitted without altering the result. Indeed, one can verify that the metric $D(\mathbf{0},\mathbf{0}) = +\infty$ and $D(x_+, x_-) = PP(x_+ - x_-)$ if $(x_+, x_-) \neq (\mathbf{0},\mathbf{0})$ satisfies COMP', MON', and ACONS'. Moreover, condition COMP' in parts (a) and (b) cannot be relaxed to consistency with $(x_+, x_-) \longmapsto CPP_0(x_+ - x_-)$ on $\{(b1_\tau, a1_0), 0 < a \leq b, \tau > 0\}$ (the set of projects with two transactions without simultaneous inflows and outflows) or replaced by consistency with $(x_+, x_-) \longmapsto CPP(x_+ - x_-)$ on $\{(x_+, x_-) \in \mathrm{X}\times\mathrm{X}\colon x_+ - x_- \in \mathrm{C}$ and there is $\tau \in \mathrm{R}_+$ such that $x_+$ vanishes on $[0,\tau)$ and $x_-$ is constant on $[\tau, +\infty)\}$ (the set of conventional projects without simultaneous inflows and outflows). There exists a rich class of functions on $\mathrm{X}\times\mathrm{X}$, besides $(x_+, x_-) \longmapsto PP(x_+ - x_-)$, that satisfy MON', ACONS', LSC', and the above-mentioned modifications of COMP'. Particular examples are $D_\infty(x_+, x_-) = \inf\{\tau \in \mathrm{R}_+\colon x_+(\tau) \geq x_-(t)\ \forall t \in \mathrm{R}_+\}$, $D_h(x_+, x_-) = \inf\{\tau \in \mathrm{R}_+\colon x_+(t) \geq x_-(t+h)\ \forall t \geq \tau\}$, $h \in \mathrm{R}_{++}$ as well as convex combinations of them. Observe that none of these examples can be represented as a function of the cumulative cash flow stream $x_+ - x_-$. The metric $D_\infty$ represents the moment in time at which the cumulative cash inflow first exceeds the total cash outflow, which is a minor modification of the payback metric mentioned in Section 1. Notably, in addition to ACONS', we have $\min\{D_\infty(x_+, x_-), D_\infty(y_+, y_-)\} \leq D_\infty(x_+ + y_+, x_- + y_-)$. That is, the mentioned earlier impossibility result in Sokolov (2024, Section 4.2) in some sense no longer holds for payback metrics on $\mathrm{X}\times\mathrm{X}$. It is a promising direction for future research to characterize all metrics on $\mathrm{X}\times\mathrm{X}$ that satisfy conditions MON', ACONS', LSC', and the above-mentioned modifications of COMP'.

## 6. Conclusion

Unlike other capital budgeting techniques such as the net present value or the internal rate of return, which can be derived from fundamental economic principles and axiomatic foundations,[6] the payback period lacks a formal characterization that would allow for systematic analysis of its properties, consistency, and domain of applicability. This paper addresses this gap by developing an axiomatization of the payback period capital budgeting technique. We propose a set of economically meaningful axioms that uniquely characterize the payback period. Our result implies that the traditional definition of payback period can be uniquely extended to all cash flows, establishing a formal basis for its application to nonconventional cash flows. This finding proves robust across a range of project space definitions. The formula for the payback period we characterize is not new, appearing in at least Hajdasiński (1993), Vilensky et al. (2002, Section 8.3), and Cheremushkin (2016). However, its existing justifications rely entirely on economic intuition—capital is fully recovered only at the time from which the cumulative cash flow stream remains nonnegative. Our result provides a formal axiomatic foundation for this intuition.

[6] An axiomatic foundation of the net present value is presented in Norberg (1990), Promislow (1994), Smolyak (2002), and Armerin (2014). For an axiomatic foundation of the internal rate of return and its modifications, see Promislow & Spring (1996), Vilensky & Smolyak (1999), and Sokolov (2023, 2024).

## 7. References

Armerin, F. (2014). An axiomatic approach to the valuation of cash flows. *Scandinavian Actuarial Journal*, *2014*(1), 32–40. https://doi.org/10.1080/03461238.2011.628408

Bhandari, S. B. (1985). Discounted payback period. *Journal of Financial Education*, *14*, 1–16. https://www.jstor.org/stable/41948078

Bhandari, S. B. (2009). Discounted payback period-some extensions. *Journal of Business and Behavioral Sciences*, *21*(1), 28–38.

Boyle, G., & Graeme, G. (2006). Payback without apology. *Accounting and Finance*, *46*, 1–10. https://doi.org/10.1111/j.1467-629X.2006.00158.x

Brealey, R. A., Myers, S. C., & Allen, F. (2020). *Principles of corporate finance*. The McGraw-Hill.

Buskes, G., & van Rooij, A. (1997). *Topological spaces: from distance to neighborhood*. Springer.

Cheremushkin, S. V. (2016). Generalized method of determining the payback period for both conventional and non-conventional cash flows: Ready-to-use excel formulas and UDF. SSRN Electronic Journal. https://doi.org/10.2139/ssrn.1982827

Fabozzi, F. J., & Peterson, P. P. (2003). *Financial management and analysis*. John Wiley & Sons.

Graham, J. R. (2022). Presidential address: corporate finance and reality. *The Journal of Finance*, *77*(4), 1975–2049. https://doi.org/10.1111/jofi.13161

Hajdasiński, M. M. (1993). The payback period as a measure of profitability and liquidity. *The Engineering Economist*, *38*(3), 177–191. https://doi.org/10.1080/00137919308903096

Jahnke, H., & Simons, D. (2008). A rationale for the payback criterion: an application of almost stochastic dominance to capital budgeting. SSRN Electronic Journal. https://dx.doi.org/10.2139/ssrn.1299584

Kim, B.-C., Shim, E., & Reinschmidt, K. F. (2013). Probability distribution of the project payback period using the equivalent cash flow decomposition. *The Engineering Economist*, *58*, 112–136. https://doi.org/10.1080/0013791X.2012.760696

Lefley, F. (1996). The payback method of investment appraisal: a review and synthesis. *International Journal of Production Economics*, *44*(3), 207–224. https://doi.org/10.1016/0925-5273(96)00022-9

Liao, M. (1976). Modified payback analysis. *Management Accounting*, *58*(3), 19–21.

Monteiro, G. A., Slavík, A., & Tvrdý, M. (2018). *Kurzweil-Stieltjes integral. Theory and applications*. World Scientific.

Narayanan, M. P. (1985). Observability and the payback criterion. *The Journal of Business*, *58*(3), 309–323. https://www.jstor.org/stable/2353000

Nijhawan, G. (2012). Measuring project profitability. In N. Tikoo (Ed.), *Project management* (pp. 168–182). Excel Books Private Limited.

Norberg, R. (1990). Payment measures, interest, and discounting: An axiomatic approach with applications to insurance. *Scandinavian Actuarial Journal*, *1*, 14–33. https://doi.org/10.1080/03461238.1990.10413870

Promislow, S. D. (1994). Axioms for the valuation of payment streams: a topological vector space approach. *Scandinavian Actuarial Journal*, *2*, 151–160. https://doi.org/10.1080/03461238.1994.10413937

Promislow, S. D. (1997). Classification of usurious loans. In M. Sherris M. (Ed.). *Proceedings of the 7th International AFIR Colloquium* (pp. 739–759). Institute of Actuaries of Australia.

Promislow, S. D., & Spring, D. (1996). Postulates for the internal rate of return of an investment project. *Journal of Mathematical Economics*, *26*(3), 325–361. https://doi.org/10.1016/0304-4068(95)00747-4

Rappaport, A. (1965). The discounted payback period. *Management Services*, *2*(4), 30–36.

Siziba, S., & Hall, J. H. (2021). The evolution of the application of capital budgeting techniques in enterprises. *Global Finance Journal*, *47*, 100504. https://doi.org/10.1016/j.gfj.2019.100504

Smolyak, S. A. (2002). *Investment project evaluation under risk and uncertainty (expected effect theory)*. Nauka. (in Russian)

Sokolov, M. V. (2023). An effective interest rate cap: a clarification. arXiv:2307.08861. https://doi.org/10.48550/arXiv.2307.08861

Sokolov, M. V. (2024). NPV, IRR, PI, PP, and DPP: a unified view. *Journal of Mathematical Economics*, *114*, 102992. https://doi.org/10.1016/j.jmateco.2024.102992

Vilensky, P. L., Livshits, V. N., & Smolyak, S. A. (2002). *Evaluation of the efficiency of investment projects. Theory and practice.* Delo. (in Russian)

Vilensky, P. L., & Smolyak, S. A. (1999). Internal rate of return of project and its modifications. *Audit and Financial Analysis*, *4*, 203–225.

Wambach, A. (2000). Payback criterion, hurdle rates and the gain of waiting. *International Review of Financial Analysis*, *9*(3), 247–258. https://doi.org/10.1016/S1057-5219(00)00028-4

Weingartner, H. M. (1969). Some new views on the payback period and capital budgeting decisions. *Management Science*, *15*(12), B594–B607. https://doi.org/10.1287/mnsc.15.12.B594

Yard, S. (2000). Developments of the payback method. *International Journal of Production Economics*, *67*(2), 155–167. https://doi.org/10.1016/S0925-5273(00)00003-7

## 8. Appendix: proofs

**Proof of Proposition 3.**

(a)⇒(b). Let $D$ satisfies COMP', MON', ACONS', and LSC'. Pick $(x_+, x_-) \in \mathrm{X} \times \mathrm{X}$.

CLAIM 1: $x_-(0) > 0 \;\&\; x_- \preccurlyeq x_+ \Rightarrow D(x_+, x_-) = 0$.

Let $x_-(0) > 0$ and $x_- \preccurlyeq x_+$. Pick $\tau \in \mathrm{R}_{++}$ and set

$$y_+(t) = \begin{cases} 0 & \text{if } t \in [0, \tau) \\ x_+(t) & \text{if } t \in [\tau, +\infty) \end{cases}, \quad y_-(t) = \begin{cases} x_-(\tau) & \text{if } t \in [0, \tau) \\ x_+(t) & \text{if } t \in [\tau, +\infty) \end{cases}. \tag{1}$$

By construction, $(y_+, y_-) \in \mathrm{X} \times \mathrm{X}$, $y_+ \preccurlyeq x_+$, $x_- \preccurlyeq y_-$, $y_+ - y_- \in \mathrm{P}_0$, and $CPP_0(y_+ - y_-) = \tau$. Hence, $D(x_+, x_-) \leq D(y_+, y_-) = CPP_0(y_+ - y_-) = \tau$, where the inequality follows from MON' and the first equality follows from COMP'. Since $\tau \in \mathrm{R}_{++}$ was arbitrary, $D(x_+, x_-) = 0$.

CLAIM 2: $x_- \preccurlyeq x_+ \Rightarrow D(x_+, x_-) = 0$.

Let $x_- \preccurlyeq x_+$. Set $y_+^{(c)}(t) = x_+(t) + c$, $y_-^{(c)}(t) = x_-(t) + c$, $c \in \mathrm{R}_{++}$. By Claim 1, for any $c \in \mathrm{R}_{++}$, $D(y_+^{(c)}, y_-^{(c)}) = 0$. Condition LSC' implies $D(x_+, x_-) = 0$.

(b)⇒(c). Let (b) hold. Set $F(x_+, x_-) = PP(x_+ - x_-)$. Put $\mathrm{P}_1 = \{x \in \mathrm{P}$: there exists $\tau \in \mathrm{R}_{++}$ such that $\sup_{t \in [0,\tau)} x(t) < 0$ and $\inf_{t \in [\tau,+\infty)} x(t) \geq 0\}$. Note that $\mathrm{P}_1 \subset \mathrm{C}$. Pick some $(x_+, x_-) \in \mathrm{X} \times \mathrm{X}$.

Claim 1: $F(x_+, x_-) \in \mathrm{R}_{++} \Rightarrow D(x_+, x_-) \leq F(x_+, x_-)$.

Set $\tau = F(x_+, x_-)$ and define $y_+$ and $y_-$ as in Eq. (1) with this $\tau$. It follows from the definition of $F(x_+, x_-)$ that $(y_+, y_-) \in \mathrm{X} \times \mathrm{X}$, $y_+ \preccurlyeq x_+$, $x_- \preccurlyeq y_-$, $y_+ - y_- \in \mathrm{P}_0$, and $CPP_0(y_+ - y_-) = \tau$. Hence, $D(x_+, x_-) \leq D(y_+, y_-) = CPP_0(y_+ - y_-) = \tau$, where the inequality follows from MON' and the first equality follows from COMP'.

Claim 2: $x_+ - x_- \in \mathrm{P}_1 \Rightarrow D(x_+, x_-) = F(x_+, x_-)$.

Let $x_+ - x_- \in \mathrm{P}_1$ and let $\tau \in \mathrm{R}_{++}$ be such that $\sup_{t \in [0,\tau)} (x_+ - x_-)(t) < 0$ and $\inf_{t \in [\tau,+\infty)} (x_+ - x_-)(t) \geq 0$. Clearly, $\tau$ is unique and $F(x_+, x_-) = \tau$. Put $c = \sup_{t \in [0,\tau)} (x_+ - x_-)(t)$ and $d = \sup_{t \in [\tau,+\infty)} (x_+ - x_-)(t)$. Since functions $x_+$ and $x_-$ are bounded, $c$ and $d$ are real numbers. As $x_+ - x_- \in \mathrm{P}_1$, we have $c < 0$ and $d \geq 0$.

Claim 1 implies that $D(x_+, x_-) \leq \tau$. In order to prove the reverse inequality, set

$$z_+(t) = \begin{cases} x_-(t) + c & \text{if } t \in [0,\tau) \\ x_-(t) + d & \text{if } t \in [\tau,+\infty) \end{cases}, \; z_- = x_-.$$

By construction, $(z_+, z_-) \in \mathrm{X} \times \mathrm{X}$, $x_+ \preccurlyeq z_+$, $z_- \preccurlyeq x_-$, $z_+ - z_- \in \mathrm{P}_0$, and $CPP_0(z_+ - z_-) = \tau$. Hence, $D(x_+, x_-) \geq D(z_+ - z_-) = CPP_0(z_+ - z_-) = \tau$, where the inequality follows from MON' and the first equality follows from COMP'.

Claim 3: $D(x_+, x_-) \geq F(x_+, x_-)$.

We have to show that $x_+(t) \geq x_-(t)$ for all $t > D(x_+, x_-)$. Assume by way of contradiction that there exists $\tau$ such that $\tau > D(x_+, x_-)$ and $x_+(\tau) < x_-(\tau)$. Set $y_+ = a1_\tau + b1_{\tau+\varepsilon}$, $y_- = a1_0$, $(a, b, \varepsilon) \in \mathrm{R}^3_{++}$. For all $(a, b, \varepsilon) \in \mathrm{R}^3_{++}$, we have $(y_+, y_-) \in \mathrm{X} \times \mathrm{X}$, $y_+ - y_- \in \mathrm{P}_1$, and $D(y_+, y_-) = \tau$ (by Claim 2). Since $x_+ - x_-$ is bounded and right-continuous at $\tau$, for sufficiently large positive $a$ and $b$ and small positive $\varepsilon$, we have $(x_+ + y_+) - (x_- - y_-) \in \mathrm{P}_1$ and $D(x_+ + y_+, x_- + y_-) = \tau + \varepsilon$ (by Claim 2). Condition ACONS' and the equalities $D(y_+, y_-) = \tau$ and $D(x_+ + y_+, x_- + y_-) = \tau + \varepsilon$ imply $D(x_+, x_-) \geq \tau + \varepsilon > \tau$, which is a contradiction.

Claim 4: $D(x_+, x_-) \leq F(x_+, x_-)$.

If $F(x_+, x_-) = +\infty$, then the inequality in the claim holds trivially. If $F(x_+, x_-) \in \mathrm{R}_{++}$, then the inequality follows from Claim 1. Finally, if $F(x_+, x_-) = 0$, then it follows from the definition of $F(x_+, x_-)$ that $x_- \preccurlyeq x_+$, and the last assumption imposed on $D$ in (b) implies $D(x_+, x_-) = 0$.

Combining Claims 3 and 4, we get $D = F$.

(c)⇒(a). Let $D(x_+, x_-) = PP(x_+ - x_-)$. Clearly, $D$ satisfies COMP', MON', and ACONS'. In order to verify LSC', note that since the topology on P is finer than the topology of pointwise convergence, for every $t \in \mathrm{R}_+$, the linear functional on $\mathrm{P} \times \mathrm{P}$ defined by $(x_+, x_-) \mapsto x_+(t) - x_-(t)$ is continuous (Buskes & van Rooij, 1997, Example 14.6(ii), p. 217). Hence, for any $d \in \mathrm{R}_{++}$, the set $\{(x_+, x_-) \in \mathrm{X} \times \mathrm{X}: D(x_+, x_-) \leq d\}$ is closed as the intersection of a closed in $\mathrm{P} \times \mathrm{P}$ set $\bigcap_{t: t > d} \{(x_+, x_-) \in \mathrm{P} \times \mathrm{P}: x_+(t) - x_-(t) \geq 0\}$ and $\mathrm{X} \times \mathrm{X}$. □

**Proof of Proposition 1.**

Set $F(x)=\inf\{\tau\in \mathrm{R}_{+}: x(t)\geq 0\ \ \forall t\geq \tau\}$ and $G(x)=\inf\{CPP_0(y): y\in \mathrm{P}_0, y\preccurlyeq x\}$. Put $\mathrm{P}_1=\{x\in \mathrm{P}: x(0)<0,\ x$ is nondecreasing, and there is $\tau\in \mathrm{R}_{++}$ such that $x(\tau)\geq 0\}$. We begin with a proof of part 3°.

3°. (a)⇒(b). Let $D$ be a payback period. It is straightforward to verify that the function on $\mathrm{X}\times\mathrm{X}$ defined by $(x_+,x_-)\longmapsto D(x_+-x_-)$ satisfies COMP', MON', and ACONS'. In view of the implication "(b)⇒(c)" in Proposition 3, it is sufficient to prove that $\mathbf{0}\preccurlyeq x\Rightarrow D(x)=0$. Assume that $\mathbf{0}\preccurlyeq x$ and pick some $\tau\in \mathrm{R}_{++}$. As $-1_0+1_\tau\preccurlyeq \mathbf{0}\preccurlyeq x$, we have $D(x)\leq D(-1_0+1_\tau)=CPP_0(-1_0+1_\tau)=\tau$, where the inequality follows from MON and the first equality follows from COMP. Since $\tau\in \mathrm{R}_{++}$ was arbitrary, we get $D(x)=0$.

(b)⇒(a). Straightforward.

(b)⇔(c). Pick $x\in \mathrm{P}$. We consider three cases.

CASE 1: $F(x)=0$.

It follows from the definition of $F(x)$ that $\mathbf{0}\preccurlyeq x$. For any $\tau\in \mathrm{R}_{++}$, we have $-1_0+1_\tau\preccurlyeq\mathbf{0}\preccurlyeq x$. Hence, $0\leq G(x)=\inf\{CPP_0(y): y\in \mathrm{P}_0, y\preccurlyeq x\}\leq \inf\{CPP_0(-1_0+1_\tau): \tau\in \mathrm{R}_{++}\}=0$.

CASE 2: $0<F(x)<+\infty$.

Set $\tau=F(x)$ and $z=-c1_0+c1_\tau$, $c\in \mathrm{R}_{++}$. It follows from the definition of $F(x)$ that $z\preccurlyeq x$ for sufficiently large $c$. Thus, $G(x)=\inf\{CPP_0(y): y\in \mathrm{P}_0, y\preccurlyeq x\}\leq CPP_0(z)=\tau$. In order to prove the reverse inequality, note that for any $\varepsilon>0$, there is $t\in(\tau-\varepsilon,\tau)\cap \mathrm{R}_{++}$ such that $x(t)<0$. This shows that $y\in \mathrm{P}_0\ \&\ y\preccurlyeq x\Rightarrow CPP_0(y)\geq\tau$. Therefore, $G(x)=\inf\{CPP_0(y): y\in \mathrm{P}_0, y\preccurlyeq x\}\geq\tau$.

CASE 3: $F(x)=+\infty$.

It follows from the definition of $F(x)$ that for any $\tau\in \mathrm{R}_{++}$, there is $t>\tau$ such that $x(t)<0$. This proves that $\{y: y\in \mathrm{P}_0, y\preccurlyeq x\}=\emptyset$ and, therefore, $G(x)=\inf\{CPP_0(y): y\in \mathrm{P}_0, y\preccurlyeq x\}=\inf\emptyset=+\infty$.

2°. The proof is similar to that of $(x_+,x_-)\longmapsto PP(x_+-x_-)$ satisfies LSC' in the implication "(c)⇒(a)" in Proposition 3.

1°. To show the independence of COMP, MON, and ACONS, we provide three examples of functions from P to $\overline{\mathrm{R}}_+$ that satisfy two of the conditions while violating the third.

OBSERVATION 1: the constant function $D(x)=0$ meets MON and ACONS while violating COMP.

Trivial.

OBSERVATION 2: the function defined by $D(x)=F(x)$ if $x\in \mathrm{P}_1$ and $D(x)=+\infty$ otherwise meets COMP and ACONS while violating MON.

Clearly, $D$ satisfies COMP. Since $\mathrm{P}_1$ is a convex cone and $F$ satisfies ACONS, so does $D$. It follows from representation (b) in part 3° that $D$ is not a payback period and, therefore, must violate MON.

OBSERVATION 3: the function $D(x)=\sup\{\tau\in \mathrm{R}_{++}: x(t)<0\ \ \forall t\in[0,\tau)\}$ (with the convention $\sup\emptyset=0$) meets all the conditions except ACONS.

Clearly, $D$ satisfies COMP and MON. It follows from representation (b) in part 3° that $D$ is not a payback period and, therefore, must violate ACONS. □

**Lemma 1.**

*Let $\alpha$ be a nonincreasing discount function, $x \in \mathrm{P}$, and $t \in \mathrm{R}_{++}$. If $x$ is nonpositive on $[0, t)$, then $x^{(\alpha)}(t) \leq \alpha(t)x(t)$. In particular, if in addition $x(t) \leq 0$ or $x(t) < 0$, then so is $x^{(\alpha)}(t)$.*

**Proof.**

Using integration by parts, we get

$$x^{(\alpha)}(t) = x(0) + \int_0^t \alpha(\tau)\mathrm{d}x(\tau) = x(0) + \alpha(t)x(t) - \alpha(0)x(0) - \int_0^t x(\tau)\mathrm{d}\alpha(\tau) \leq \alpha(t)x(t),$$

where we use that $\alpha(0) = 1$, $\alpha$ is nonincreasing, and $x$ is nonpositive on $[0, t)$. □

**Proof of Proposition 2.**

To each $D: \mathrm{P} \to \overline{\mathrm{R}}_+$ assign the function $D^{(\alpha)}: \mathrm{P} \to \overline{\mathrm{R}}_+$ given by $D^{(\alpha)}(x) = D(x^{(\alpha)})$. It is straightforward to verify that the map $D \longmapsto D^{(\alpha)}$ defines a bijection between the set of payback periods and the set of discounted payback periods. The inverse map is given by $D \longmapsto D^{(1/\alpha)}$. Now the statement of the proposition follows from Proposition 1. □